\documentclass{aa}

\usepackage{graphicx}
\usepackage{txfonts}

\begin{document}

\title{The oldest tidally induced bar-like galaxy in the IllustrisTNG cluster}

\author{Ewa L. {\L}okas
}

\institute{Nicolaus Copernicus Astronomical Center, Polish Academy of Sciences,
Bartycka 18, 00-716 Warsaw, Poland\\
\email{lokas@camk.edu.pl}}

\abstract{
New JWST observations have  revealed the presence of a significant number of high-redshift barred galaxies. The origin
of these bars remains unclear, and their properties appear difficult to reconcile with the results of cosmological
simulations of galaxy formation. I present an example of a tidally induced bar-like galaxy formed at $z = 2.9$ in the
TNG100 suite of the IllustrisTNG simulations. The galaxy was identified among the sample of bar-like galaxies studied
before and has the earliest bar formation time among the tidally induced subsample of those objects. Its disk
transformed into a bar as a result of a close interaction with a massive progenitor of a brightest cluster galaxy
(BCG). It remained on a tight orbit around the host and survived until the present, losing most of its initial mass and
becoming red but preserving its prolate shape. Even before the interaction, at $z=3.5$, the galaxy experienced a few
mergers, which elongated its shape. This temporary distortion also made it look like a bar with spiral extensions of up
to 6 kpc. The long-lived bar formed later was about 3 kpc long and grew over the next few gigayears. This example
demonstrates that high-$z$ bars should not be sought among the progenitors of present-day simulated barred galaxies but
rather among the tidally interacting early population of galaxies in forming groups and clusters. Some of these
galaxies may have survived as ellipticals, and some may have merged with their BCGs. }


\maketitle

\section{Introduction}

\nolinenumbers

The most interesting recent discoveries made with JWST observations include the identification of well-developed barred
galaxies at high redshifts \citep{Guo2023, Costantin2023, Amvrosiadis2025}. These new data also allowed unprecedented
determinations of the bar fraction in the Universe at early times. It has been demonstrated that the bar fraction
decreases with redshift but does not become negligible, even at $z = 3$ \citep{LeConte2024, Guo2025, Geron2025}. On the
theoretical side, cosmological simulations experience some difficulty in reproducing the observed dependence of bar
fraction on time and in producing mature bars at such early times in cosmic evolution \citep{Peschken2019, Reddish2022}.
Although bars can be identified inside disks even at $z = 4$, for example in IllustrisTNG50 \citep{Rosas2022}, they
tend to be small, not exceeding the length of 1 kpc. In the zoom-in simulations of \citet{Bi2022}, the
high-redshift bars were also of sub-kiloparsec lengths and gas-rich.

The bars observed at high redshifts, for example the Milky Way-like bar at $z = 3$ discovered by \citet{Costantin2023}
with JWST or the barred spiral galaxy found at $z = 4.4$ by \citet{Tsukui2024} using ALMA data, both with lengths of
about 3 kpc, did not necessarily evolve into anything similar to the Milky Way bar by the present time. In fact, the
analysis of the evolution of the bars in the Milky Way-M31 analogues from TNG50 \citep{Pillepich2024} shows that their
bars grew mostly via disk instability from small elongations, unlike the mature bars observed by JWST. An alternative
way to form a bar, and a mature one almost at once, is via interactions with other objects \citep{Noguchi1987,
Gerin1990, Miwa1998, Berentzen2004, Lokas2014, Lokas2016, Lokas2018}. Such interactions could involve both flybys and
mergers, which are much more frequent in the early stages of galaxy evolution.

In this paper I present such a scenario for one galaxy in a sample of bar-like galaxies from the IllustrisTNG project
studied in \cite{Lokas2021}. These galaxies were identified at $z = 0$ among sufficiently resolved and prolate systems.
I studied the evolutionary histories of these galaxies and divided them into classes depending on their formation
scenarios and properties. One of the classes (class A) included objects whose bar-like structures had been induced by
an interaction with a more massive companion. The galaxy described here had the earliest bar formation time of this
class, $t = 2.2$ Gyr ($z = 2.9$).

\section{Formation and properties of the bar-like galaxy}

The object of interest for this study comes from the TNG100-1 suite of simulations from the IllustrisTNG project
\citep{Springel2018, Marinacci2018, Naiman2018, Nelson2018, Pillepich2018}, which follow the evolution of dark matter
and baryons in a box of size 100 Mpc, solving for gravity and hydrodynamics and applying additional prescriptions for
processes like star formation and feedback. The galaxy was selected from the publicly available simulation data
described by \citet{Nelson2019} as a subhalo with identification number ID96556 at the last simulation output ($z=0$).
It was found among sufficiently massive galaxies with stellar masses $M_* > 10^{10}$ M$_\odot$ and strongly prolate
systems in which the intermediate-to-long axis ratio of the stellar component is $b/a < 0.6$. The axis ratios were
estimated from the eigenvalues of the mass tensor of the stellar component included within two stellar half-mass radii,
$2 r_{1/2}$. Among the 277 bar-like galaxies identified in this way, 77 were found to be formed via tidal interaction
with a more massive companion. This tidally induced subsample is described in more detail in \citet{Lokas2025}.

\begin{figure}
\centering
\includegraphics[width=9cm]{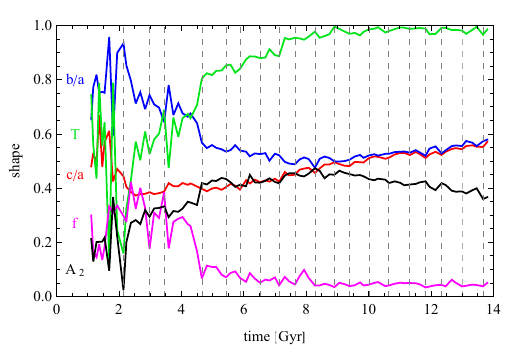}
\caption{Evolution of different measures of shape over time for galaxy ID96556. The lines show the axis ratios
$b/a$ (blue) and $c/a$ (red), the triaxiality parameter $T$ (green), the rotation parameter $f$ (magenta), and the bar
mode $A_2$ (black). Vertical dashed lines indicate pericenter passages around the most massive progenitor of galaxy
cluster ID96500.}
\label{shape}
\end{figure}

Figure~\ref{shape} illustrates the evolution of different measures of the shape of the galaxy over time, with the
intermediate-to-long, $b/a$, and short-to-long, $c/a$, axis ratios of the stellar component shown with the blue and red
line, respectively. The green line plots the triaxiality parameter $T = [1-(b/a)^2]/[1-(c/a)^2]$, which is below 1/3
for oblate and above 2/3 for prolate systems. The rotation parameter $f$ (magenta line) is related to the fraction of
stars on circular orbits and measures the amount of rotational support in the system \citep{Genel2015}, with $f \sim
0.4$ corresponding to well-developed disks. The most important parameter for this analysis is the commonly used measure
of the bar strength  \citep{Athanassoula2002} in the form of the $m=2$ mode of the Fourier decomposition of the surface
density distribution of stellar particles projected along the short axis. It is given by $A_2 (R) = | \Sigma_j m_j
\exp(2 i \theta_j) |/\Sigma_j m_j$, where $\theta_j$ is the azimuthal angle of the $j$th star, $m_j$ is its mass, and
the sum is over all particles in a given radial bin. The single-value measurements presented here (black line) were
obtained using all stars within two stellar half-mass radii, $2 r_{1/2}$.

All the parameters shown in Fig.~\ref{shape} indicate that at $t = 2.2$ Gyr ($z=2.9$) the galaxy's morphology underwent
a dramatic and permanent change: the axis ratio $b/a$ decreased and the triaxiality $T$ and the bar mode $A_2$
increased, signifying the formation of a bar. In particular, the latter rapidly crossed the threshold value of 0.2,
which can be adopted as a signature of strong bar formation, and remained above it until the end of the evolution. The
inspection of the galaxy's environment at that time reveals that the transformation coincided with the first strong
interaction with massive companions in a forming protocluster, ID96500 (marked with the first vertical dashed line in
Fig.~\ref{shape}).

The interaction also caused a strong mass loss in the galaxy. Before the encounter, at $t = 1.8$ Gyr, the galaxy
reached its maximum mass, $6.7 \times 10^{11}$ M$_\odot$, of which 87\% was dark matter, 11\% was gas, and 2\% was
stars. Immediately after the first interaction, the galaxy started to lose dark matter and gas, though the stars
continued to form as a result of a short starburst induced by the interaction. As the galaxy continued on its orbit,
the gas was lost at the third pericenter passage and the stars also started to get stripped. At the end of the
evolution, the mass of the galaxy was only about $2 \times 10^{10}$ M$_\odot$, half of which was dark matter and the
other half was in the stellar component, and the galaxy stellar population became red.

At the time of the first interaction, $t = 2.2$ Gyr ($z=2.9$), the protocluster contained three galaxies more massive
than ID96556 (one of which was more distant and infalling). The galaxy ID96556 interacted first with the most massive
of the three, which at that time had a mass of almost $7 \times 10^{12}$ M$_\odot$, passing it at a pericenter distance
of only 34 kpc on a prograde orbit (with an angle of 45 deg between the internal and orbital angular momenta just
before the interaction). Soon after, it also interacted with the other two galaxies, but much more weakly. The three
companions later merged to form the brightest cluster galaxy (BCG). The vertical dashed lines in Fig.~\ref{shape}
indicate the pericenter passages around the most massive companion and the whole BCG later on.

Interestingly, even before the first interaction, at $t = 1.8$ Gyr ($z=3.5$) the galaxy experienced a temporary
increase in the triaxiality $T$ and the bar mode $A_2$, as well as a decrease in the axis ratio $b/a$. This means that
it underwent a bar-like distortion already then. The inspection of its interaction history at that time shows that it
experienced a number of mergers with smaller satellites. To be precise, at $t = 1.3$ Gyr ($z=4.7$) there were six
subhalos more massive than $10^9$ M$_\odot$ in its vicinity that merged with it between $t = 1.5$ and $2.0$ Gyr,
including one as massive as $2 \times 10^{10}$ M$_\odot$. The mergers must have distorted the morphology, causing the
elongation.

\begin{figure*}
\centering
\includegraphics[width=6.3cm]{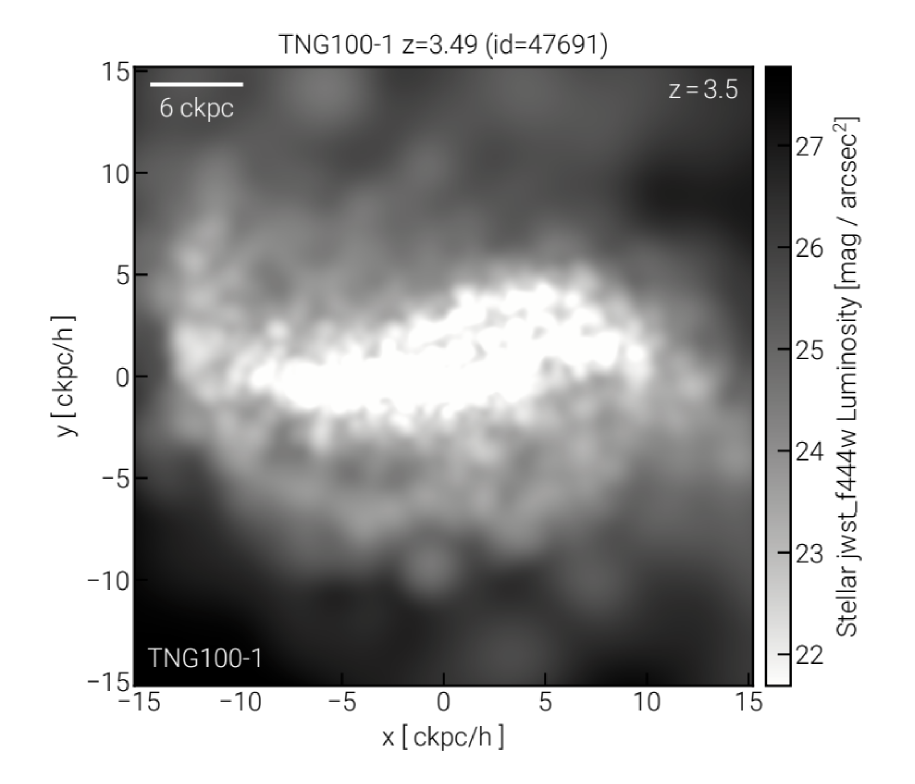}
\hspace{-0.5cm}
\includegraphics[width=6.3cm]{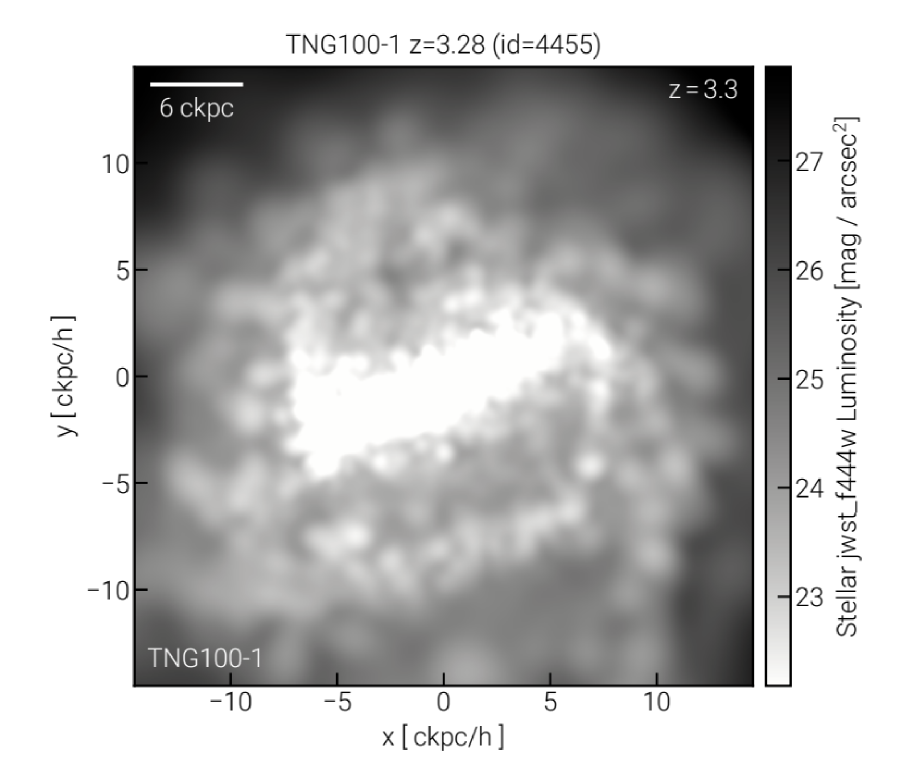}
\hspace{-0.5cm}
\includegraphics[width=6.3cm]{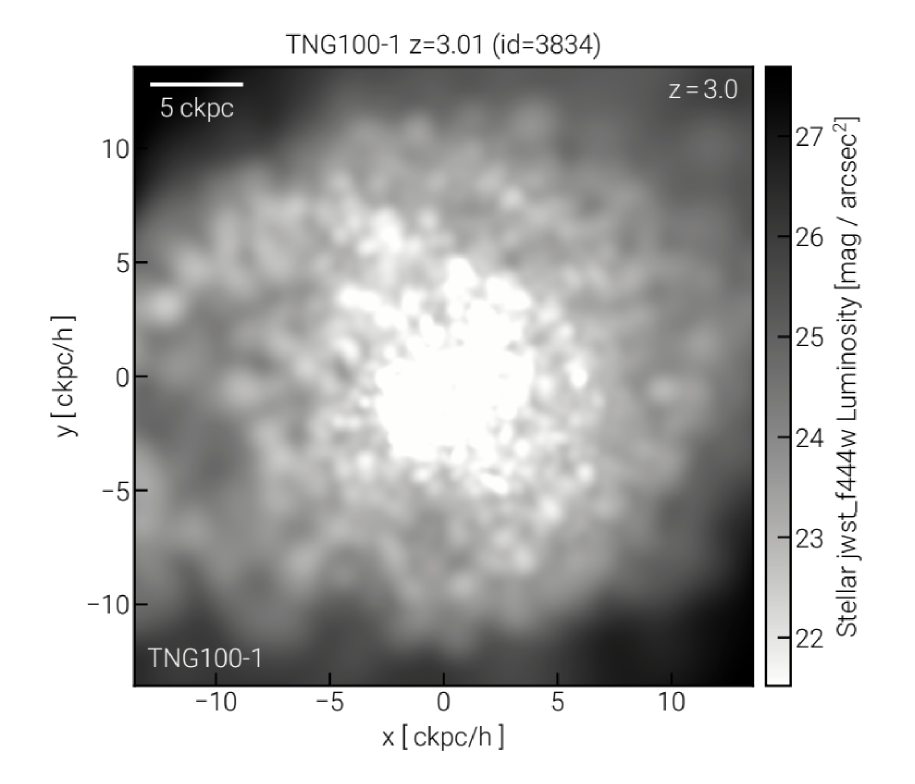}  \\
\includegraphics[width=6.3cm]{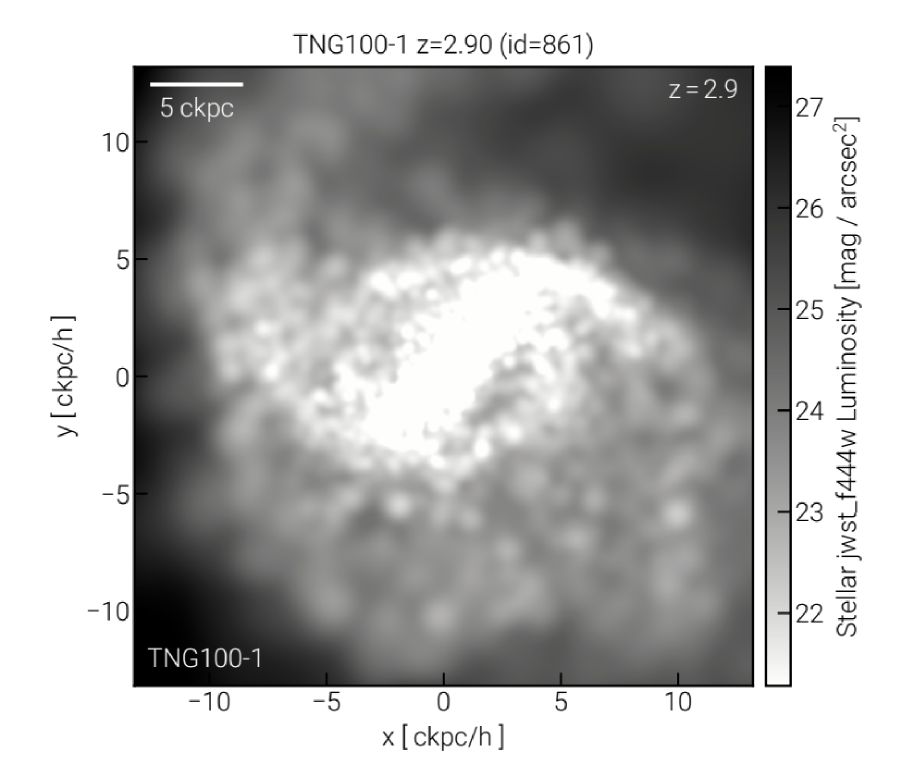}
\hspace{-0.5cm}
\includegraphics[width=6.3cm]{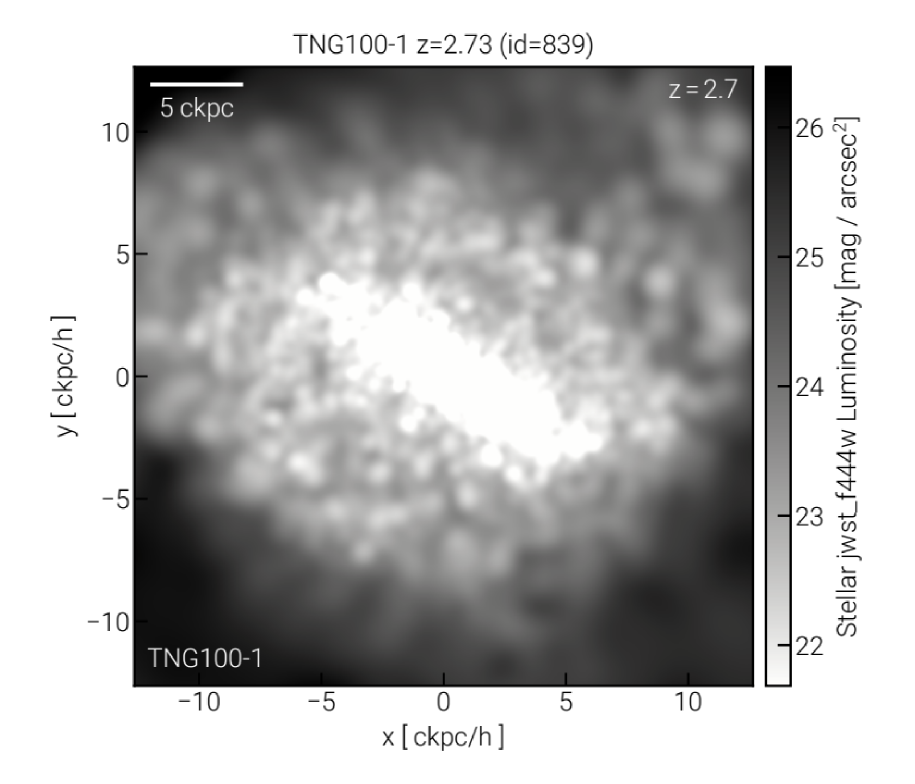}
\hspace{-0.5cm}
\includegraphics[width=6.3cm]{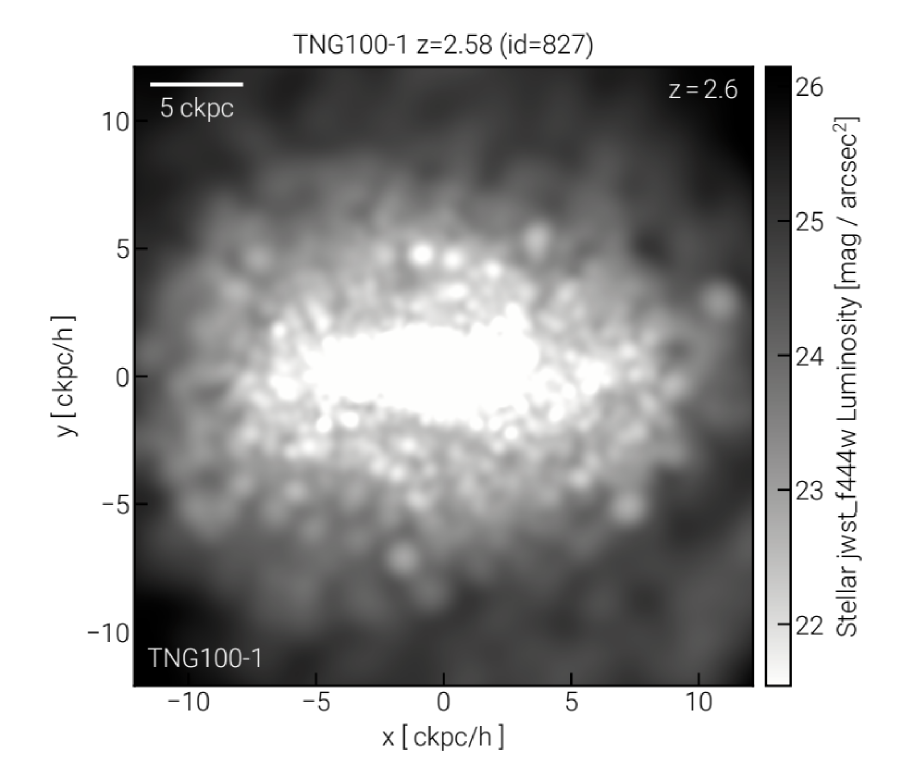}
\caption{Face-on images of the galaxy ID96556 at six subsequent simulation outputs from $t = 1.8$ to $t=2.5$ Gyr
($z=3.5-2.6$) as they would be observed in the F444W filter of JWST.}
\label{images}
\end{figure*}

Figure~\ref{images} shows images of the stellar component of the galaxy ID96556 as it would be seen in the F444W filter
of JWST in a face-on projection (along the shortest axis of the stellar component). The six panels correspond to the
simulation outputs from $t = 1.8$ to $t=2.5$ Gyr ($z=3.5-2.6$). The upper-left image shows the elongation created by
the mergers, which survived until the next snapshot (upper-middle panel) but disappeared later (upper-right panel). The
lower row of panels shows the bar-like shapes created as a result of the interaction with the massive progenitor of the
BCG, ID96500. This time, the bar was a permanent feature and survived until the end of the evolution. As shown in
Fig.~\ref{shape}, at the fourth pericenter passage ($t=4.7$ Gyr) the galaxy's bar strength increased further and
its rotation nearly ceased.

\begin{figure}
\centering
\includegraphics[width=7.5cm]{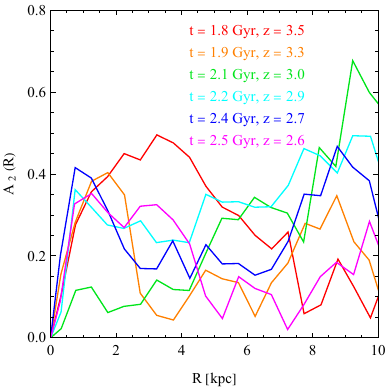}
\caption{Profiles of bar mode $A_2 (R)$ for galaxy ID96556 at different times.}
\label{a2profiles}
\end{figure}

Detailed properties of the bar at different stages of the evolution can be read from the profiles of the bar mode $A_2
(R)$ shown in Fig.~\ref{a2profiles}; the measurements were carried out in bins of $\Delta R = 0.5$ kpc of the
cylindrical radius. The profiles are plotted for different times, corresponding to the images shown in
Fig.~\ref{images}. For most snapshots, the profiles show a characteristic bar mode shape, which increases with radius,
reaches a maximum, and then decreases. The only discrepant profile is the one at $t=2.1$ Gyr ($z=3.0$, green line),
which increases with radius at larger distances from the galaxy center. At this time the galaxy is at the pericenter
and undergoes strong tidal distortion by the more massive companion. For later outputs, the feature disappears and the
bar mode returns to its normal behavior.

The length of the bar at different stages can be estimated as the radius $R$ at which $A_2(R)$ drops to half its
maximum value. For $t=1.8$ Gyr ($z=3.5$, red line) the $A_2$ profile is very extended and the bar seems to be
incredibly long, on the order of 6.3 kpc. In this case, the phase of the bar mode is varying slowly with radius in the
outer parts, signifying a transition to tidally induced spiral arms. In the next snapshot, at $t=1.9$ Gyr ($z=3.3$,
orange line), its length is already much shorter, around 2.5 kpc. A bar of similar length is recreated at $t=2.4$ Gyr
($z=2.7$, blue line), after the interaction at the pericenter.

\begin{figure}
\centering
\hspace{0.4cm}
\includegraphics[width=3.5cm]{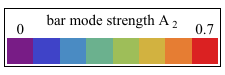}\\
\vspace{-0.2cm}
\includegraphics[width=9cm]{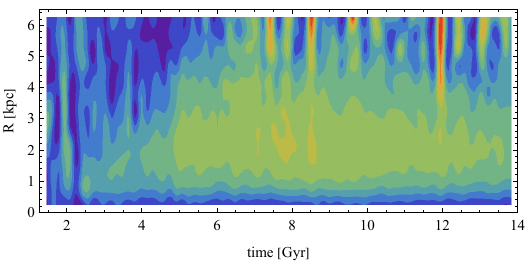}
\caption{Evolution of the profile of the bar mode $A_2 (R)$ over time for galaxy ID96556.}
\label{a2modestime}
\end{figure}

The whole evolution of the $A_2(R)$ profiles can be seen in Fig.~\ref{a2modestime}, where they are plotted in a
color-coded form as a function of time. The plot allows the evolution of both the strength and the length of the bar to
be followed in detail. For example, it can be seen that after the fourth pericenter passage, at 4.7 Gyr, the bar became
not only stronger, but also longer. The bar was at its strongest around 8 Gyr, and which point it started to weaken as
a result of tidal stirring at multiple passages around the BCG. The redder stripes at larger radii indicate temporary
elongations at the pericenters.

\section{Discussion}

I have presented a "proof-of-concept" example of an old bar-like galaxy formed via tidal interaction with a massive
companion, a progenitor of the BCG of a forming cluster. The galaxy is unique in the sense that it was found to
be the oldest among the bar-like galaxies formed this way (the second oldest appeared 1.5 Gyr later)  in the TNG100-1
run of IllustrisTNG that were previously identified in the last simulation snapshot, which corresponds to the present
time \citep{Lokas2021, Lokas2025}. This means that in spite of multiple interactions with the cluster BCG, the object
survived until the present.

This channel of the formation of high-redshift bars seems probable since tidal interactions as well as mergers occur
more frequently at the early stages of galaxy evolution. It is even possible that high-$z$ bars preferentially form via
tidal interactions, also because the galactic disks are then hotter, more gas-rich, and more turbulent, which makes the
onset of the inherent bar instability more difficult. In fact, many of the high-$z$ bars studied by \citet{Guo2023,
Guo2025} turn out to have companions. In particular, at $z>1.5$ the percentage of barred galaxies with companions
appears to be much higher than for unbarred disks.

The configuration in which the bar was induced in the galaxy studied here is to some extent similar to the one found by
\citet{Smail2023} for their observed barred galaxy at $z>4$. The real object also seems to belong to a group of
galaxies, although in their case it is the main galaxy that appears to have a bar-like feature. If this structure is of
tidal origin, it must have been induced by a smaller companion or mergers.

The simulated bar-like galaxy presented here may be exceptional in the sense that it survived until the present time.
At present, however, it would not look like a typical barred galaxy with a bar embedded in a disk. It would rather
appear as a red, weakly rotating elliptical, although quite elongated. This suggests that one should not look for early
bars in simulations among the progenitors of present-day barred disks or Milky Way-M31 analogues \citep{Rosas2022,
Pillepich2024}. Most of these galaxies seem to have formed their bars during a long, secular evolution via dynamical
instability of cold enough, massive disks, and grew their bars gradually, from a very low initial strength and a short
length.

In this study, I used the TNG100 simulation of the IllustrisTNG suite, which has a larger simulation box at the cost of
a lower resolution compared to TNG50. This results in the smaller bars not being resolved in TNG100 (unlike in TNG50)
and the impression that TNG50 produces overall smaller bars. This should not be an issue since in this work I am
interested in long bars seen by JWST, and those should be resolved in TNG100. Instead, finding configurations similar
to the one studied here can be problematic in TNG50 because this simulation contains only one massive galaxy cluster.
Indeed, as discussed in \citet{Lokas2024}, one can find significantly fewer bar-like galaxies in TNG50 than expected.

As mentioned above, flybys and mergers in controlled as well as cosmological simulations tend to produce long, mature
bars soon after the interaction \citep{Lokas2018, Lokas2022} rather than grow them from small perturbations. This is
confirmed by the recent study of bars in the suite of Auriga simulations of Milky Way-like galaxies by
\citet{Fragkoudi2025}, who find that bars formed at higher redshifts (up to $z=2$) tend to be born long via
interactions, while those forming later start shorter and grow in time. On the other hand, using high-resolution
zoom-in simulations of disk galaxies, \citet{Bi2022} found them to have only short, sub-kiloparsec bars at high
redshifts in spite of the fact that all their bars at $z > 2$ formed via interactions. This difference can be traced to
the specific setup of their simulations, which did not include active galactic nucleus feedback. The lack of this
feature is known to prevent or slow the formation of bars because of the strong accumulation of gas in the galaxy
center \citep{Athanassoula2016, Bonoli2016, Rosas2025}. This interpretation is supported by the fact that the small
early bars of \citet{Bi2022} are very gas-rich.

Further examples of early bars in cosmological simulations can be identified by studying interactions and mergers at
early times. Many such bars do not survive until the present, not only in terms of morphology but even as separate
galaxies. An example of such an occurrence is provided by the very configuration studied here. One of the massive
members in the ID96500 group also had a bar induced as a result of an interaction with the most massive progenitor of
this BCG. However, neither the bar nor the galaxy survived because the latter soon merged with the BCG. Such examples
of sizable bar-like galaxies formed early via interactions in IllustrisTNG will be discussed in detail in a follow-up
paper.

\begin{acknowledgements}
I am grateful to the anonymous referee for useful comments and to the IllustrisTNG team for making their
simulations publicly available. Computations for this work have been performed using the computer cluster at the
Nicolaus Copernicus Astronomical Center of the Polish Academy of Sciences (CAMK PAN).
\end{acknowledgements}

\end{document}